\documentclass[sigconf]{acmart}
\usepackage{booktabs}
\usepackage{graphics}
\usepackage[ruled]{algorithm2e} 
\usepackage{enumitem}
\usepackage{tabularx}
\usepackage{subcaption}
\usepackage{url}
\usepackage[skip=3pt]{caption}
\usepackage{xspace}
\usepackage{bbm}

\newcommand{\libname}{\textit{Lumos}\xspace}
\copyrightyear{2020} 
\acmYear{2020} 
\setcopyright{acmlicensed}
\acmConference[KDD '20]{Proceedings of the 26th ACM SIGKDD Conference on Knowledge Discovery and Data Mining}{August 23--27, 2020}{Virtual Event, CA, USA}
\acmBooktitle{Proceedings of the 26th ACM SIGKDD Conference on Knowledge Discovery and Data Mining (KDD '20), August 23--27, 2020, Virtual Event, CA, USA}
\acmPrice{15.00}
\acmDOI{10.1145/3394486.3403306}
\acmISBN{978-1-4503-7998-4/20/08}

\begin{document}
\fancyhead{} 
\title{Lumos: A Library for Diagnosing Metric Regressions in Web-Scale Applications}
\author{Jamie Pool, Ebrahim Beyrami, Vishak Gopal, Ashkan Aazami, Jayant Gupchup, Jeff Rowland, Binlong Li, Pritesh Kanani, Ross Cutler, and Johannes Gehrke}
\affiliation{\institution{ Microsoft}\city{Redmond} \state{WA} }
\email{{jamie.pool, ebbeyram, vishak.gopal, ashkana, jayant.gupchup, jeff.rowland, bil, prkanani, ross.cutler, johannes}@microsoft.com}

\renewcommand{\shortauthors}{Pool, et al.}

\begin{abstract}
Web-scale applications can ship code on a daily to weekly cadence. These applications rely on online metrics to monitor the health of new releases. Regressions in metric values need to be detected and diagnosed as early as possible to reduce the disruption to users and product owners. Regressions in metrics can surface due to a variety of reasons: genuine product regressions, changes in user population and bias due to telemetry loss (or processing) are among the common causes. Diagnosing the cause of these metric regressions is costly for engineering teams as they need to invest time in finding the root cause of the issue as soon as possible. We present  \libname, a Python library built using the principles of A/B testing to systematically diagnose metric regressions to automate such analysis. \libname has been deployed across the component teams in Microsoft's Real-Time Communication (RTC) applications  \emph{Skype} and \emph{Microsoft Teams}. It has enabled engineering teams to detect 100s of real changes in metrics and reject 1000s of false alarms detected by anomaly detectors. The application of \libname has resulted in freeing up as much as $95\%$ of the time allocated to metric-based investigations. In this work, we open source \libname and present our results from applying it to two different components within the RTC group over millions of sessions. This general library can be coupled with any production system to manage the volume of alerting efficiently.

\end{abstract}


\maketitle

\section{Introduction}
\label{s:intro}
The health of online services is monitored by tracking Key Performance Indicator (KPI) metrics over time. Regressions in these KPIs require a follow up as they could be indicative of service misbehavior
resulting in costs and the potential of loss of customers. However, it is time-consuming and costly to track down the root cause of every KPI regression. A single anomaly can often take days to weeks to investigate thoroughly. In this paper, we address the two core difficulties described above. The first issue is that we need to have a \emph{trustworthy} anomaly detector; by trustworthy we mean an anomaly detector that only cries wolf if there is something wrong in the service. The second issue is that we need to quickly determine the root cause of a KPI regression given thousands of hypotheses. We propose a novel methodology that encompasses existing, domain-specific anomaly detectors, but has been found to reduce the false-positive alert rate by over $90\%$ in our deployments. Also, our methodology provides insight into locating the root cause of a KPI incident. 

What are the challenges in creating this methodology? The first major challenge in using even state-of-the-art anomaly detectors is their high false-positive rate \cite{ren2019, vallis2014, luminol}.  The issue arises when the metric is non-stationary in terms of time and user demographics. In online monitoring systems distinguishing between system failures and regressions due to change in external factors (i.e., usage pattern) is critical. Regressions based on external factors are not necessarily actionable as they may not be under our control. For the Skype and Teams applications, we track hundreds of different metrics.  We slice our data across dozens of different dimensions, such as tenant, geographical location, and platform.  This results in thousands of time-series datasets that are monitored for anomalous behavior. The volume of tracked metrics means we experience hundreds of KPI alerts per month. After ruling out regressions due to external factors, we are left with a more manageable number of system/service failures (such as a code change, a new experiment, or an outage) which are actionable from an engineering point of view. Having now isolated the real anomalies, there could be many reasons for the anomaly, and thus it was our second major challenge to find ways to help engineers to investigate the root cause of such anomalies.

Additionally, any proposed solution needs to be able to run at scale.  The process needs to be automated to run at the cadence of the anomaly detection algorithm. Since we have a large call volume, any proposed solution needs to be able to handle anomalies that show small changes in metric value.   The solution needs to be easily picked up by new engineering teams to incorporate into their anomaly detection workflow. The output results need to be easily interpreted by engineers who do not have a data science or statistics background.

The main idea in the development of \libname was to include domain knowledge in the full workflow for investigations. The first key point we identified is that there are different population groups who we expect to have different behaviors. For instance, calls coming from regions with connectivity issues would experience more dropped calls. When investigating an increase in call drop rate, we would find more calls coming from these regions. Here the change in the metric could be explained by a change in the user population.  To handle this problem at scale, we identified a set of the population identifying features that we don't expect to vary over time. We introduce a process of checking the data for bias with respect to these features and normalizing the results.   To do this we use  A/B testing principles.  %The second piece of domain knowledge we can utilize is for the scenario where we determined that a population change wasn't the root cause.  In this case, we have thousands of telemetry features per call that could provide an explanation to an engineer of a potential issue to address.
In scenarios where the regression is not explained by bias, telemetry data is used to provide insight to the engineering teams. The library has a component that creates a prioritized list of those features.   This part of the process directly needs the expertise of the engineering teams who can interpret the telemetry features.

\libname is a library for investigating regressions in metrics based on binary variables. It eliminates the very expensive manual process of establishing whether a change is due to a shift in population or the introduction of a product change.  \libname also provides a prioritized list of the most important variables in explaining the change in the metric value. Skype and Teams have been using \libname and the performance has shown to be very effective at both filtering out the anomalies due to population bias and determining the root cause of the alerts. While \libname is primarily being used in anomaly detection scenarios, it serves the wider purpose of understanding the difference in a metric between any two datasets. It compares a control and treatment dataset and is agnostic to the time series component.  For example, \libname can be used to determine why one tenant had more call drops than another during the same time period. We have also used the library for other projects where there is a need to identify biases, or explaining a difference between two datasets.

The key contributions of this paper are as follows:
\begin{itemize}
    \item We present a workflow for investigating anomalies that reduce the number of false-positives, and we give a simple algorithm for prioritizing features that describe the anomaly. (Section \ref{s:lumos})
    \item We show the results from a real-world deployment of \libname in Microsoft RTC applications and describe in detail several illustrative case studies. (Section \ref{s:deployment})
    \item We share lessons learned based on several years of experience for the benefit of other practitioners. (Section \ref{s:lessons})
\end{itemize}

The library \libname is open sourced and available for use on GitHub \cite{lumos}.  

We continue with a discussion of related work in the next section.

\section{Related Work}
\label{s:relwork}
Monitoring KPIs for anomalous behavior is a common business practice.   State of the art algorithms such as from Twitter \cite{vallis2014} and LinkedIn \cite{luminol}, give an unsupervised approach to capturing anomalies. More recently a supervised anomaly detection algorithm was introduced by Microsoft in \cite{ren2019} where a neural network approach was introduced using ideas from visual saliency research.  

All anomaly detection algorithms we have tested have relatively high false-positive rates, and our goal is to reduce this false positive rate using principals from randomized controlled experiments (A/B testing).  There has been extensive literature in randomized controlled experiments, e.g., \cite{kohavi2007, crook2009, kohavi2012, deng2013, deng2018, xie2018, lee2018}. A/B testing has been extensively used in experimentation as a way to control for bias and determine if product changes impact a given metric.  In our scenario of tracking dashboard metrics, we can assign a control dataset containing data from calls before the anomaly and a treatment dataset from calls during the anomaly.  However, there are no controls for the population biases between the two datasets.  This formulation is in line with ideas from the literature for causal inference in observational studies.  This has long been a topic of research, e.g., \cite{Imbens2015, Rubin1974}.  In an observational study, the treatment group is self-selected, and the validity of any results needs to be disentangled from any biases caused by the selection process.  Wherein A/B tests directly control for the biases during the design of the experiment.

One technique often employed in observational studies is propensity score matching. The idea was introduced by Rubin and Rosenbaum in \cite{rosenbaum1983}, where the propensity score is the conditional probability of a sample being in the control or treatment group based on a set of covariates. The matching refers to the process of locating samples in the two datasets with similar scores. Once data is matched, the two datasets have comparable populations with respect to the set of covariates. This process has been shown to remove the bias between the two datasets.  There are multiple types of matching in propensity score matching; we were motivated by a caliper-based matching procedure introduced in  \cite{cochran1973} and multiple authors \cite{Rosenbaum1985, dehejia2002} have investigated combining the caliper match with the propensity scores.  Their work has shown this approach to be effective in removing bias from the datasets. An appropriate choice of caliper width is the standard deviation of the propensity scores times a caliper coefficient as shown in \cite{austin2011}. 

Many tools can be used to give a ranking of feature importance for machine learning algorithms.  These vary from being specific to a given algorithm, like out of bag scores for random forests, to interpreting more generic black-box models.  This is an extensive area of research; see \cite{lundberg2017, lundberg2020,ribeiro2016, saabas} for more information.  However, there are a couple of difficulties with adopting this approach.  The first is that you have to build a model on each dataset that accurately predicts the outcome metric. We are working with data with correlated features and high-class imbalance.  It is likely that without careful tuning that a model would not learn the underlying patterns in the data. This tuning would differ for every incident we fed the tool.  There are automated machine learning pipelines \cite{elsken2019neural, olson2016evaluation, jin2019auto} that cover the process of taking raw data and finding an appropriate machine learning model, which can overcome this initial challenge. The second challenge is comparing feature importance between the two models. In our experience building models and comparing feature rankings was not sufficient to point engineers in the correct direction to resolve the metric regression. The primary difficulty is that understanding features correlated with a change in a metric is a different task than finding features correlated with the metric itself. There are additional techniques for discovering relationships in large datasets such as association rule learning \cite{agrawal1993, webb2000}.   These models will have the same challenge of comparing two different sets of relationships.   %This analysis is outside of the scope of what these tools were built to handle.  

Other previous works on root cause analysis for anomalies include \cite{lin2016} where the authors look to understand changes in the volume of reported issues. They construct a pruning algorithm that efficiently identifies the effective combination of slicers from a subset of tens of features that determine the location of the anomaly. This work is not directly applicable to our situation as we are looking at KPI metrics, which have different properties than the count of issues reported, and the scale of our hypothesis features is in the hundreds. In \cite{raff2016}  they develop an algorithm for identifying the key differences between two datasets with respect to a given metric. This work provides a methodology for running a statistical test between the two datasets to determine feature importance and ranking. A key difference between this approach and ours is the necessary component of checking for bias and using this to reduce the high false-positive anomaly rate. In \cite{tolomei2017interpretable}, the authors address extracting interpretation (insights) from tree-based ensemble classification methods.
The idea is to take all the true negative labels and transform decision boundaries until they become true positive. This transformation provides actionable insights to the consumers of the model on what is required to get more positive outcomes. This work does not address the questions of extracting bias conditions or factors that explain metric deltas.

\section{Lumos}
\label{s:lumos}
\setlength{\belowcaptionskip}{-5pt}
\begin{figure*}[t]
    \centering
        \includegraphics[width=.7\textwidth]{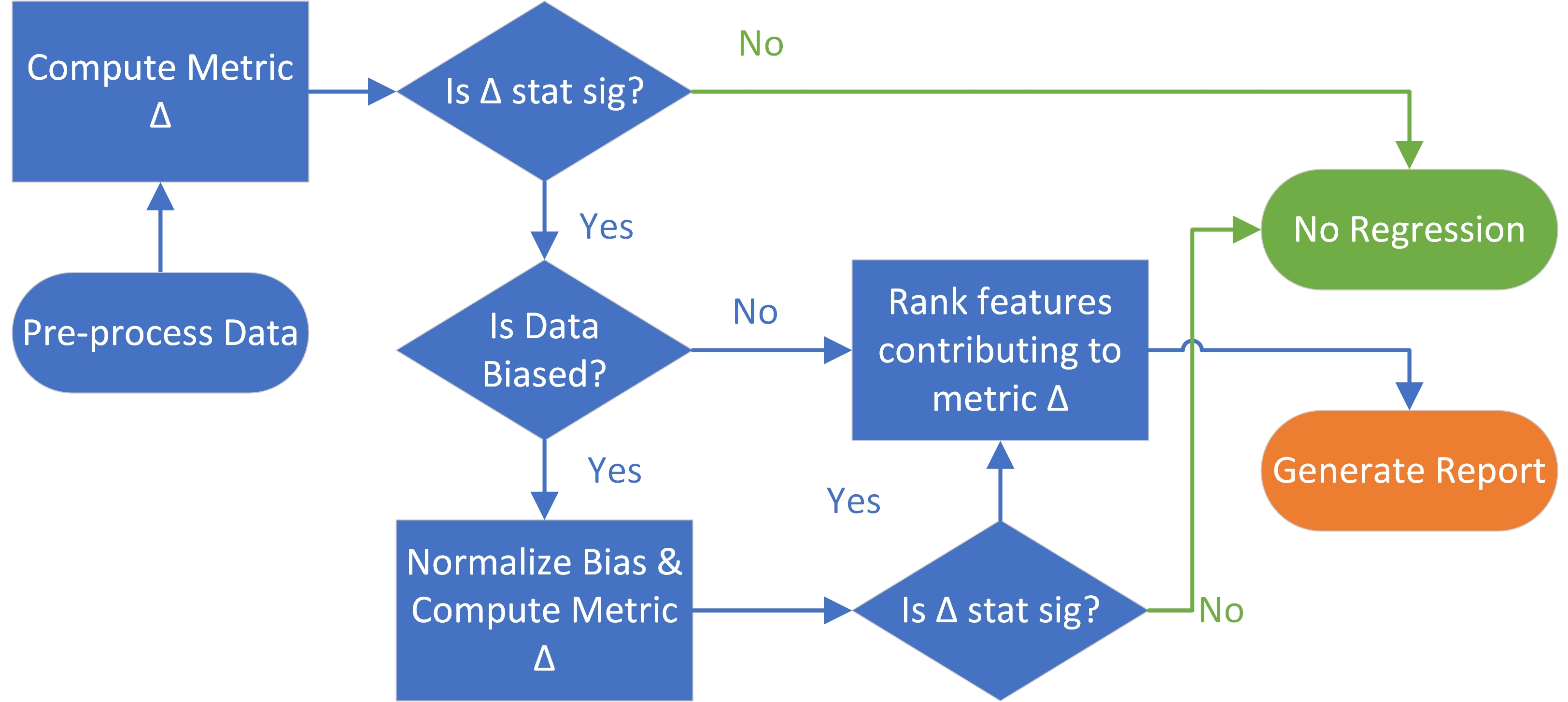}
    \caption{\libname Workflow}
    \label{fig:lumos_flow}
\end{figure*}
\libname is a library designed for diagnosing metric regressions. The primary workflow for metric diagnostics inputs two datasets and a configuration file.  \libname  uses the principles of A/B testing and compares the dataset before the metric anomaly (i.e., control dataset), and the dataset during the metric anomaly (i.e., treatment dataset). Each dataset is a tabular dataset where each row indicates a single sample, and the column values include the metrics of interest. In particular, these columns include: the variable representing the KPI (i.e., \emph{Target metric}), variables describing the population (i.e., \emph{Invariant features}), and variables that provide hypotheses for diagnosing the regression in the metric (i.e., \emph{Hypothesis features}). The configuration file contains hyper-parameters for running the workflow and details which columns in the datasets correspond to the metric, invariant, and hypothesis columns.  

The end-to-end flow for using \libname is shown in Figure \ref{fig:lumos_flow} and Algorithm \ref{alg:Lumos}. \libname will begin by verifying if the regression in the metric between these two datasets is statistically significant. It will then follow up with a population bias check and bias normalization to account for any population changes between the two datasets.   If there is no statistically significant regression in the metric after the data is normalized, then the regression in the metric can be explained by the change in the population. If the delta in the metric is statistically significant, the features are ranked according to their contribution to the delta in the target metric. 

\setlength{\belowcaptionskip}{-10pt}
\begin{algorithm}[t]
\SetAlgoLined
 $(C, T) \leftarrow$ Load control, treatment datasets\\
 $(\Delta, p\text{-value}) \leftarrow$ Compare $(C,T)$ using $\chi^2$-test\\
  \eIf{$p$ is below P-value threshold}
    {
    \eIf{ there is any population bias}
        {$(C_n,T_n) \leftarrow$  Normalize datasets\\
        $(\Delta_n, p_n) \leftarrow$ Compare $(C_n,T_n)$ using $\chi^2$-test\\
        \eIf{$p_n$ is below the P-value threshold}
            {Rank features based on their contribution to $\Delta$}
            {Report population bias as the root cause of regression} 
        }
        {
        Rank features based on their contribution to $\Delta$
        }
   }
   {Report there is no statistical significant difference between $C$ and $T$}
 \caption{\libname}
 \label{alg:Lumos}
\end{algorithm}

The following sections will detail the individual components of the algorithm. \libname is implemented in a way to allow each of the individual components to be used in isolation without the full workflow. We found this useful in the situation where we only need a single component like a bias check.
Most of the work in running the tool is in the data collection. Once you have your datasets and the configuration file, the tool has a simple API to analyze the root cause of the regression.
\subsection{Data Collection}
\label{s:lumos_data}
We borrow the terminology of control and treatment datasets from controlled experimentation. The \emph{control dataset} contains data points from before the metric regression, and the \emph{treatment dataset} has data points during the metric regression. Our goal is to identify the root cause of the anomaly which is to find out which features in our dataset explain the delta between the two datasets.
%Each dataset is a tabular dataset where each row indicates a single sample, and the column values indicate the values of different features. 
There are three different sets of features that are shared between two datasets:
\begin{itemize}
    \item \emph{Target features/metrics}: the metric of interest where the anomaly is observed.  Many of our KPI metrics are based on binary outcome data as such our open source code only supports metric of this form. To extend the component of the algorithm that filters false-positives to continuous metrics is a straightforward extension achieved by replacing the statistical tests with their continuous versions. 
    \item \emph{Invariant features}: These features describe the sample population, and are used for the bias check and bias normalization.  The invariant features provide a way to have a fair comparison between the two datasets. This list should only include segments of the population that are not expected to change between the control and treatment datasets. The platform, device type, network type, and country are examples of invariant features that we do not expect to change between datasets. An example of a feature that might be expected to change would be the product version where different distributions of the version would be expected immediately after a new release. These invariant features need to be selected with domain knowledge. From an algorithmic complexity standpoint, it is not advised to include more than $~10$ of these features. The algorithm currently is designed to only handle categorical data types.
    \item \emph{Hypothesis features}: Theses features are used for analyzing the root cause of the observed regression. These columns provide a hypothesis for explaining the difference in metric values between the two datasets. This can be any additional information you collect; specifically, both raw and processed telemetry fields can serve as hypothesis variables.   In practice, more entries will result in noisier results, so we recommend having a curated list containing no more than 200 entries.  These columns can contain categorical or numerical datatypes.
\end{itemize}

The datasets that we obtain should reflect the anomaly detected in the system. For example, if the target metric increased from $5\%$ to $15\%$ then the control dataset should have the target metric at $5\%$, and the treatment dataset should have the target metric at $15\%$. For running \libname at scale, we don't manually feed the data for every case; instead, we set up a set of rules for selecting the datasets. For our Teams scenario, we have a large weekly seasonality where we expect a decrease in call volume on the weekends, and the overall network quality changes as fewer users are in corporate office settings. For these reasons, it's not fair to compare a weekday call with a weekend call. We select our treatment group to be the data during the day that our metric spiked, and the control dataset to be from the previous four weeks on the same day of the week. Since the inputs to \libname are a control and treatment dataset, it does not know how the data were selected.  To obtain the best performance, data should be chosen with respect to the time scales and seasonality effects that are present in the metric being analyzed. In our scenario, due to the total volume of calls in Skype/Teams, we randomly down-sample our datasets with the rule of thumb that to detect a $1\%$ regression in a metric we require 100K data points in each of our datasets. One approach for determining sample size to detect an effect of a given size with certain confidence is to use a power analysis; for a more in-depth discussion on this topic see \cite{ellis2010}.

\subsection{Preprocessing}
\label{s:lumos_preprocess}

The inputs to the tool include two datasets with a configuration file.  The tool initializes a preprocessing workflow, and it verifies the format of inputs is as expected. As mentioned earlier \libname currently only supports binary target metrics.

In the preprocessing step, we check to make sure all the necessary hyperparameters are defined, features defined in the configuration file are available in both datasets, and features are of the supported types. 
Next, we process the hypothesis features by:
\begin{enumerate}
\item Removing the features that contain the same entry for all samples in both datasets.
\item Reducing the number of distinct values by binning the tail values into an \emph{other} bin; the maximum number of bins can be configured by setting the corresponding hyperparameters in the configuration file.
\item Removing categorical features that contain no information on the target metric by running a chi-squared test. 
\item \emph{One-hot encoding} the categorical features.
\item Adding a new \emph{is-null} feature for each feature that indicates whether the feature value is null or not.
\end{enumerate} 
There is an option to exclude is-null features, however, we advise caution using this as missing values have meaning. In the case of telemetry data, it might indicate that a portion of the production code did not run, which would provide insights to engineers locating the root cause of the regression.

\subsection{Metric Comparison}
\label{s:lumos_metric}
A chi-squared test compares the average target metric between two datasets. %; however, a test of proportions would also be appropriate here.  
 \libname outputs a table that summarizes the difference in the metric values between the two datasets. It includes the average target metric in both control and treatment along with the resulting p-value from the statistical test and an indicator of whether the delta is statistically significant with respect to the threshold provided in the configuration file. This check is run before and after bias normalization. 
If the result of the statistical test changes from significant to insignificant after the bias normalization, then the regression in the target metric can be attributed to the population change of the invariant features.

\subsection{Bias Check and bias normalization}
\label{s:lumos_bias}
\textbf{Bias Check:}
For each invariant feature listed in the configuration file, \libname runs a chi-square test to see if the distribution of the feature is different between the two datasets. As part of the chi-squared test, the observed values are used to determine expected values under the assumption that the two datasets come from the same distribution. These values are used to compute a percent deviation for each feature.
%as
%\begin{equation*}
%    \displaystyle\max_{\substack{i\in \text{feature value}\\ j\in\text{\{Control, Treatment\}}}}\frac{|O_{ij}-E_{ij}|}{\sum_k O_{kj}}
%\end{equation*}
Invariant features with statistically significant changes in the distribution and percent deviation bigger than the threshold set in the configuration file are normalized in the bias normalization step. 

The results of the bias check are returned in a table that details the changes in user demographics.  These results provide valuable information to product owners on how their user base is changing. For example, when we had an increase in poor reliability metrics, the tool detected an increase in call volume from a specific country that explained the change in the metric rate. This information allowed product owners to prioritize future work improving the system.  

Bias normalization is done after this step to diagnose any code-related service regressions.

\begin{figure}[t]
    \begin{center}
    \includegraphics[width=0.47\textwidth]{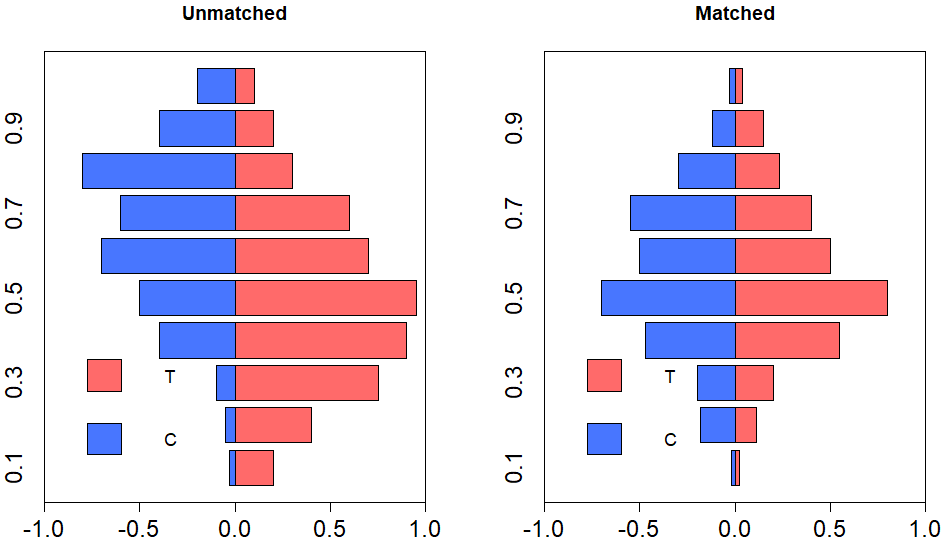}
    \end{center}
    \caption{Propensity score matching to normalize population bias}
    \label{fig:propensity}
\end{figure}

\textbf{Bias Normalization:} The features returned in the bias check step are normalized using a propensity score matching procedure. Our procedure is detailed in Algorithm \ref{alg:propensity}. A random forest model is used to predict the conditional probability that a data point is in the control or treatment dataset. The predictions from this model on the control and treatment dataset are the propensity scores. The idea is that similar feature combination will have similar propensity scores, and this score can be used to match similar samples in both datasets. The matching is done based on a histogram binning as illustrated in Figure \ref{fig:propensity}, where we use a bin width of the form \emph{caliper coefficient} times the standard deviation of the propensity scores. The caliper coefficient can be tuned in the configuration file. Reducing the caliper coefficient will enforce a stricter match between the datasets. The two histograms are matched on the intersection, which has been sufficient for our use case; however, \emph{bootstrapping techniques} \cite{pan2015,ho2007} can be added to guarantee the starting distribution of either the control or treatment dataset.

\setlength{\belowcaptionskip}{-10pt}
\newcommand{\fit}[1]{\operatorname{fit}(#1)}

\setlength{\belowcaptionskip}{-10pt}
\begin{algorithm}[t]
\SetAlgoLined
$(C, T, Invariants) \leftarrow$ control, treatment and invariant features\\
%$Data\leftarrow(C\cup T) $ Adding a column for the group label\\
$RF\leftarrow \fit{Data, X = Invariants, Y = \mathbbm{1}_{C}}$ \\
$(C_{score}, T_{score})\leftarrow \left(RF(C),RF(T)\right)$ \\
%$BinSize\leftarrow$ coefficient $*std(C_{score}\cup T_{score})$\\
$(C_{Bin}, T_{Bin})\leftarrow Bin(C_{score}, T_{score})$ \\
$(C_{n}, T_{n})\leftarrow$ Randomly match on intersection of $(C_{Bin}, T_{Bin})$
 \caption{Propensity Score Matching Procedure}
 \label{alg:propensity}
\end{algorithm}

After normalizing the data, a second bias check is run. If the bias check finds that there is still bias in the results, a warning is returned. Based on the result, the normalization can be manually restarted with hyperparameters adjusted to enforce a stricter match or include more features.

\begin{figure}[t]
    \begin{center}
        \centering
        \begin{subfigure}[b]{0.49\linewidth}
            \includegraphics[width=0.99\textwidth]{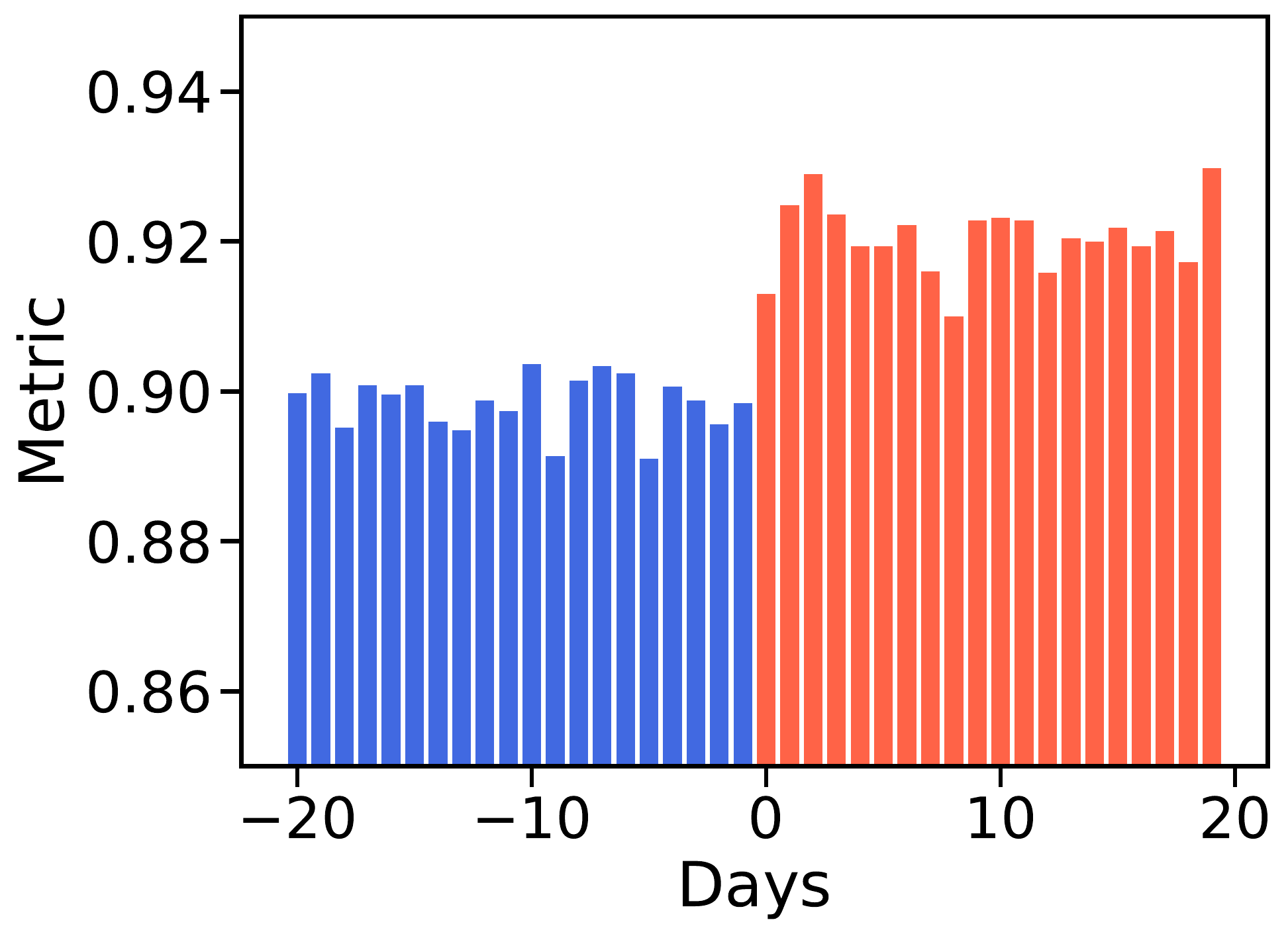}
    %        \label{fig:numbits}
        \end{subfigure}
        \begin{subfigure}[b]{0.49\linewidth}
            \includegraphics[width=0.99\textwidth]{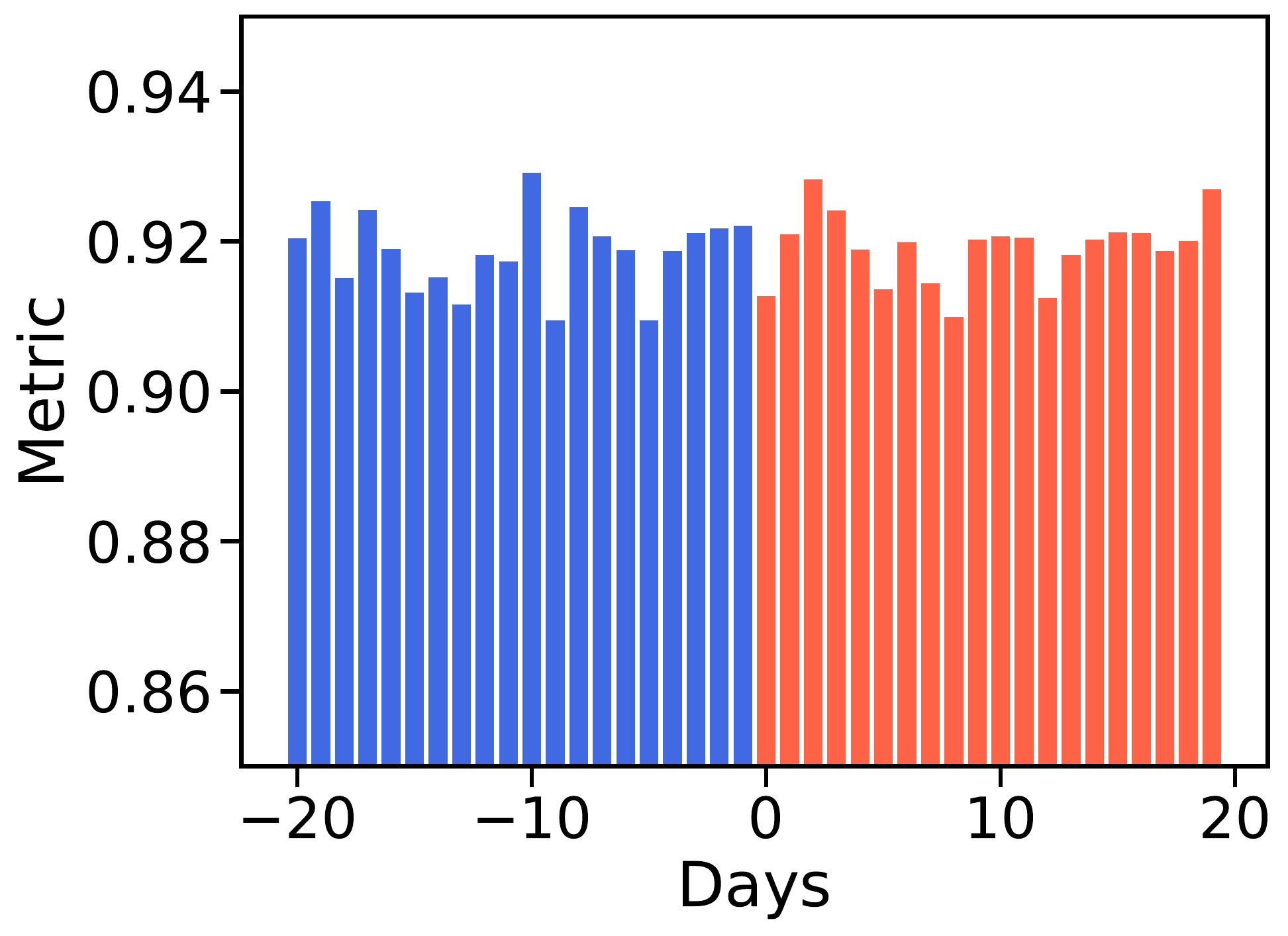}
    %        \label{fig:quadratic}
        \end{subfigure}
    \caption{Left: example of a metric observed prior to normalization. Right: after bias normalization.}
    \label{fig:bias_norm}
    \end{center}
\end{figure}

In Figure \ref{fig:bias_norm} we show an example representative of how the metric value can change after taking into account the population bias. The average of the target metric in treatment (i.e., during the anomaly) is higher than in control, but after the bias normalization, the target metric difference between treatment and initial forecast in control is not statistically significant.
\setlength{\belowcaptionskip}{-10pt}

\subsection{False-Positives}
%\jg{look over}
\libname filters the alerts for the anomaly detector into two categories.  Type-S anomalies are those that are related to systemic issues and call for an engineering intervention. Type-B anomalies are due to a bias coming from differences in populations that prevent a fair comparison of metric values.  Bias can come from population changes as accounted for in the normalization step, as well as seasonality effects that are directly accounted for in data selection. These also include cases where there is insufficient call volume for the event to be statistically significant.  In our context, Type-B anomalies are defined as false-positives alerts as they can be attributed to a population bias that the tool has accounted for.

\subsection{Feature Ranking}
\label{s:lumos_ranking}
For alerts that are not labeled as a false-positive, we provide a prioritized ranking of the hypothesis features.  During the data preprocessing step all hypothesis features were encoded so that each feature has a binary value. A univariate feature ranking algorithm is defined to determine the importance of each feature in explaining the regression in the target metric. On the normalized datasets for each feature, we determine the joint probability of feature occurrence and target metric occurrence. If the difference in probabilities is statistically significant, we consider the feature in the ranking; otherwise, we drop it. A table is constructed with a row for each feature containing: the number of failures in the treatment dataset, the number of expected failures in the treatment dataset based on the distribution of the control dataset, and the absolute and percent difference in these values.   These differences give a summary of the impact a given feature has on the metric.  Through using the tool, we found that features that explain the regression in the metric had high scores for both these metrics. To improve the ordering, an auxiliary metric that we call the \emph{hazard score} was introduced to rank the features. This score defined as the difference between the control and treatment datasets of the conditional probability of feature occurrence given that a failure occurred.  In practice using the hazard score, the features that are most relevant to diagnosing the regression have been sorted to the top of the table.  For our binary metrics, we are primarily interested when the regression is in the direction of the metric increase, and by default, this is the direction the results are sorted.  Reversing the ordering of the scores will cause the top contributors to decrease in metric value.

\section{Real-World Deployments}
\label{s:deployment}
Today we monitor hundreds of metrics related to the reliability of calling, meetings, and public switched telephone network (PSTN) services at Microsoft. Although each of the services has its internal monitoring on service-specific metrics, we use \libname as the primary tool for end-to-end (E2E) scenario monitoring across various call corridors. 

\subsection{Running at Scale}
\label{s:deply_scale}
\begin{figure}[t]
  \centering
      \includegraphics[width =0.45\textwidth]{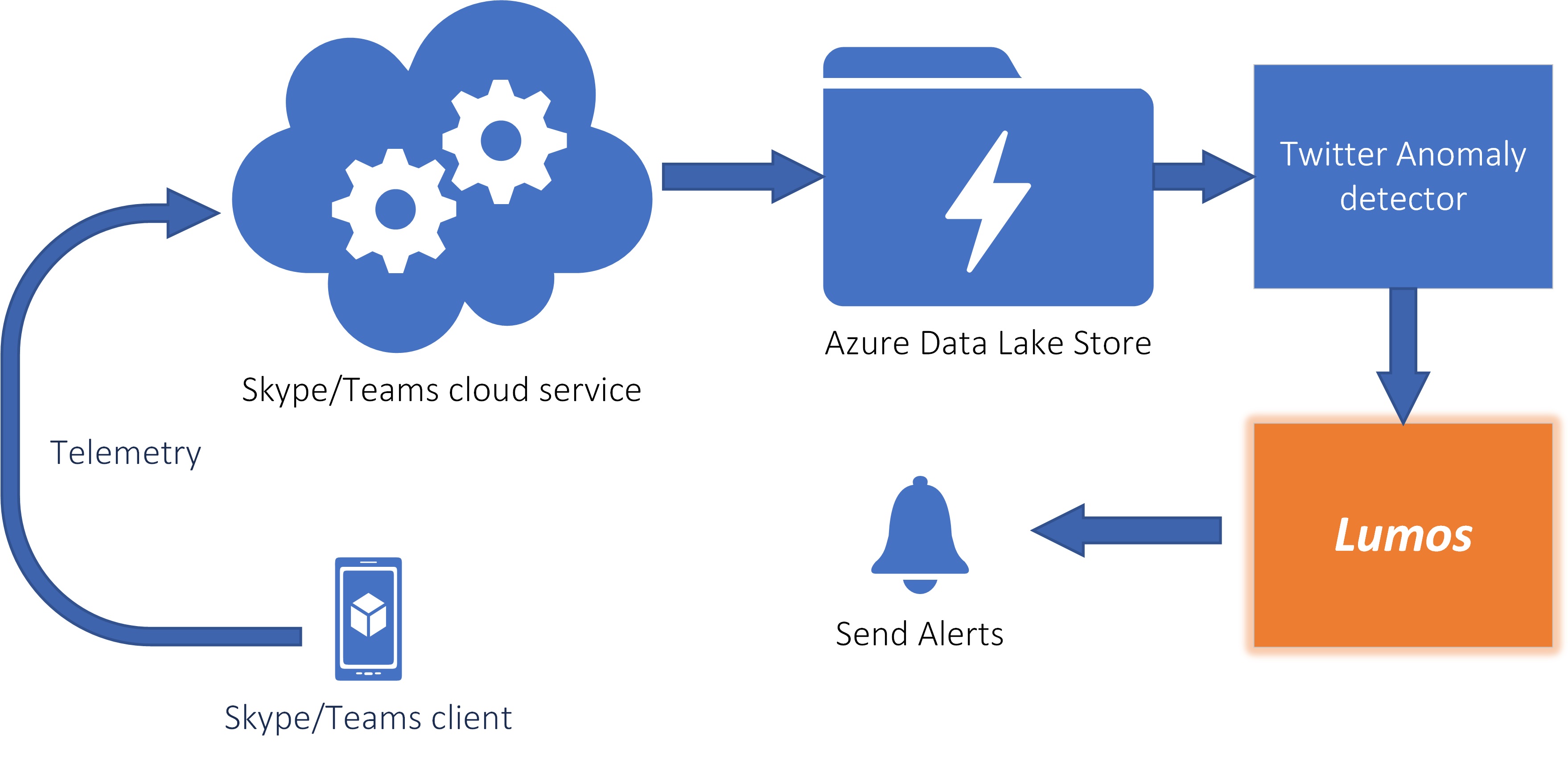}
  \caption{\libname in production}
  \label{fig:azure_data}
\end{figure}

The data flow used for applying \libname in our production dev-ops environment is shown in Figure \ref{fig:azure_data}. All the raw unstructured data required to compute the metrics for E2E scenarios resides in Azure. The raw data from services is enriched and curated to create two types of datasets:
\begin{enumerate}
    \item Aggregated time series datasets where the files feed into Azure Analysis Service to render Power-BI reports. They are also used to run the Twitter anomaly detection algorithm \cite{vallis2014} whose output feeds into \libname as a trigger.
    \item Structured datasets containing columns detailing the metric, feature, and invariant columns as corresponding to the configuration file for the metric. 
\end{enumerate}

Azure Databricks is used for running \libname and the anomaly detection algorithms, which serve as our distributed computing system. There are multiple jobs configured based on priority, complexity, and type of metrics being monitored.  Priority is set based on our knowledge of the relative importance of each metric along with the expected focus from the engineering teams. Jobs complete in Azure Analytics asynchronously. Whenever an anomaly is detected, it triggers the \libname workflow. If \libname determines it to be a legitimate issue, an incident alert (ticket) will be raised.

\setlength{\belowcaptionskip}{-5pt}

Let's take an example of monitoring the leg reliability metrics. We have 15 primary metrics each of which are being monitored against key dimensions like platform, tenant, meeting type, resulting in thousands of aggregated time series we track for a single metric. We have millions of call legs per day and each leg generates hundreds of telemetry fields serving as the input for \libname.

\subsection{False-Positive Reduction and Time Savings}
We computed the false-positive rates for KPIs from two teams and found that over $90\%$ of alerts (on a scale of hundreds of incidents per month) should not have been raised. This translated into freeing up time allocated for anomaly investigation by $65\%$ for one team and $95\%$ for the second team. In this section, we will present the methodology and validation in obtaining these results.

To compute the reduction in alerting we compared the number of alerts from our anomaly detector against the number of alerts raised by \libname.  We collected and aggregated this data over a four month period. Table \ref{tab:Alert_Reduction} displays these results for a subset of our metrics that are representative of our overall findings. The number of alerts that \libname filters out as false-positives ranges from  $90\%$ to $98\%$ based on the individual metric.  

\setlength{\belowcaptionskip}{2pt}
\begin{table}[t]
    \centering
    \scalebox{.8}{
    \begin{tabular}{l  | c }
         Metric  & Alert Reduction\\
         \hline
         Call Drops&  $95.2\%$\\
         Media Offer Failures & $98.2\%$\\
         Signaling Failures & $89.3\%$\\
        % \shortstack{Answer Seizure Ratio \\ASR} & PSTN & 93.4\%\\
        Answer Seizure Ratio  & $93.4\%$\\
        Network Effectiveness Ratio  & $94.8\%$
    \end{tabular}
    }
    \caption{The percent of alerts from the anomaly detection service that \libname filtered out.}
    \label{tab:Alert_Reduction}
\end{table}

For our analysis, we are using the Twitter Anomaly Detector \cite{vallis2014} which is set up in our deployment pipeline.  We looked into using the more recent Microsoft Anomaly Detector \cite{ren2019} by applying it to a subset of our metrics.  The alerting rate from the two anomaly detectors was comparable.  The Microsoft anomaly detector does have additional functionality to label individual points and use them in making future predictions. This feature is useful, but it does not scale to the thousands of time-series metrics we track. Without additional filtering, the volume of alerts leads to reducing trustworthiness and inefficient utilization of time.  The value \libname is providing is to using domain knowledge to filter these results into anomalies that can be explained by bias, and those with systemic causes.  Anomaly detectors are unable to make this distinction, as the relevant information is not present in the time series data. 

The time that engineers, program managers, and data analysts spend varies per incident. Some of the incidents that can be linked to experiments in flight can be root caused quickly if the results are monitored proactively. Cases where the root causing requires identifying patterns in raw telemetry or logs can take anywhere between a couple of hours by a few engineers to days (even for very senior and experienced engineers). To provide a conservative estimate for the cost savings, we assume the time to root cause is $4$ engineering hours per incident.  When there are multiple daily alerts they are grouped together for investigation, so for our estimation, we assume only $25\%$ of anomaly detector incidents are investigated. We normalize these results based on the number of people per team responsible for investigating metric regressions and find this tool frees up $65\%$ and $95\%$ of the time resources for the two teams, respectively.

To validate our results, we used a collection of known incidents where there was a regression in our service. These incidents are initially identified through avenues such as an alert raised through a tenant, adhoc debugging of code, threshold-based monitors on various metrics, and specific flighting of features with corresponding monitoring of the metrics.  Through this process, we isolated approximately 100 known incidents with established ground truth and verified that \libname correctly identified all of them for further investigation.  There were no false positives in this set of incidents.

\subsection{Case Study: Poor Call Rate} 
%\jg{rename to something more? We already covered Feature Ranking in 3.5.}
RTC applications track user-perceived quality-of-experience in the form of a call quality feedback survey \cite{gupchup2017}.  Participants are invited to rank the call on a scale from $1$ to $5$. These scores are aggregated with responses of $1$, $2$ representing poor calls. Monitoring the average value of poor call rate (PCR) allows the detection of potential changes in our service.  It is an important metric since it tracks the overall health of our media subsystem, and changes to any component of the service can show up in this metric.

One incident that \libname was able to detect for PCR involved a bug in the code that impacted video-based screen sharing  (VBSS).  Two different teams released updates and those conflicted with each other.  As a result, when users tried to use the screen sharing functionality, they experienced errors. This impacted the users' calling experience and they rated the call poorly. 

The problem was noticed with an increase in the PCR time series data.  The control and treatment datasets for \libname were selected as described in Section \ref{s:lumos_data}. The analysis from \libname identified an increase in PCR was associated with calls using the screen sharing functionality, through a feature that indicates if screen sharing is used during the call. Another important piece of information that surfaced was the primary deployment ring being impacted by this bug. This is useful information as it provides a point of context to when the changes in the code where made, reducing the time towards finding the specific issue. 

\subsection{Case Study: Adhoc Analysis}
 In addition to running \libname in a production pipeline as detailed in Section \ref{s:deply_scale}, it is also used for adhoc investigations. An example of this was at the start of the academic year the dashboards for PCR displayed increased in poor calls associated with one of our large university tenants.  Our data science team was asked to perform a manual investigation of this incident.
 
 The selection of control and treatment groups was different since there is an innate population bias in comparing calls before and after the start of the academic year, where there is an expected change in volume and characteristics of calling. To address this, we used the additional information from the dashboards that PCR was higher for adhoc meetings than scheduled meetings.  This allowed us to compare data during the academic year selecting the same dates for the treatment group composed of adhoc meetings and the control group of scheduled meetings.
 
 \libname identified a strong population bias between the two user groups, showing adhoc meetings have significantly more calls being dialed out and fewer calls being joined. The library sorts the population bins by their contribution to the bias, and the top features from this investigation are displayed in Table \ref{tab:edu_bias}.  The normalized data showed a statistically significant increase in PCR persisted between the two groups. The top features in the prioritize feature ranking pointed to connectivity issues.  These results were delivered in the hands of domain experts and they immediately identified it as  ``nudge failures''.  These failures occur when a user invites someone to join their call and there is no response from the far end. The appropriate teams responsible for this behavior were promptly notified with the information.  
 
\setlength{\belowcaptionskip}{0pt}
\begin{table}[t]
    \centering
    \begin{tabular}{c|c|c|c}
    \hline
    Feature & Bin & \shortstack{Occurrence in \\ Scheduled} & \shortstack{Occurrence in \\ Adhoc}\\
    \hline
    Leg Type &Join & $85\%$ &$31\%$\\
    Leg Type & Dial Out &$3\%$ &$33\%$\\
    Call Type & Type 1 & $47\%$ & $21\%$\\
    Leg Type & Create Call & $13\%$ & $36\%$ \\
    \end{tabular}
    \caption{Top population bias results for one of our large university tenants.}
    \label{tab:edu_bias}
\end{table}
 
 This incident highlights how you can directly select data while taking into account known biases. \libname can take these control and treatment datasets and provide the analysis.  The main impact on the business is that it provides actionable diagnostic information leading to massive savings of time and attention.

\subsection{Case Study: Non-Binary Metrics}
\label{s:deply_nonbinary}
PSTN is the international telephone system where calls are placed and audio travels through wires carrying the analog voice data between two endpoints. Customers can place and receive PSTN calls through our infrastructure. The call data travels through a network of switches as determined by a carrier external to Microsoft. We collect telemetry values through our interface but these will exclude carrier-specific telemetry. When monitoring our KPI metrics it is important to identify if there is an issue with our end of the service, or if there is a device error in the switch network.  An important metric we track to monitor the health of PSTN is the average call duration. 

\begin{figure}[t]
    \centering
    \includegraphics[width=0.45\textwidth, trim={0 0 0 1cm},clip]{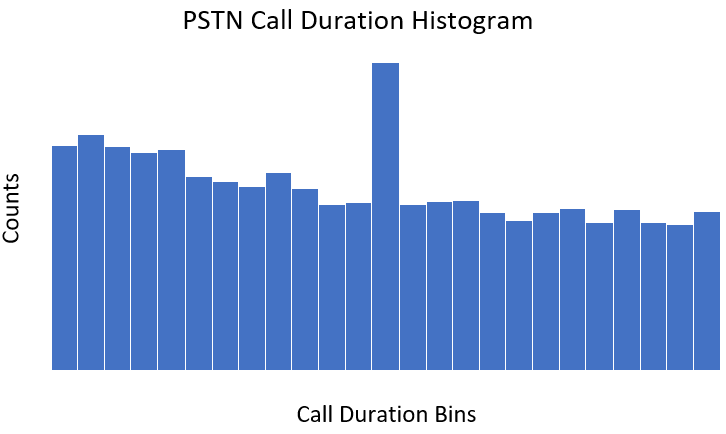}
    \caption{Histogram of call duration containing an unexpected increase in counts in the middle bin.  }
    \label{fig:CallDurationHistogram}
\end{figure}

Anomalous behavior was identified in this metric. During exploratory data analysis, it was noted that one of the bins in the call duration histogram had an unexpected increase in counts during the metric regression as displayed in Figure \ref{fig:CallDurationHistogram} that was not present in prior data. Using this information, an auxiliary binary metric was defined as an indicator function identifying if a call ended during this time bin. This allowed the problem to be transformed into the related question 'why is there an increase in calls concluding in bin $N$?' which is of the appropriate form to address with \libname.  

\setlength{\belowcaptionskip}{-2pt}

 \libname ran on the corresponding data and the analysis identified an issue for outgoing conference calls associated with a range of phone number area codes.  An investigation was conducted, and the area codes were used to narrow down the search to a set of carriers and primary tenants being impacted by the issue. This resulted in timely notification to a partner who followed up with downstream carriers. A faulty session broker device was identified as causing calls to end prematurely and was fixed.
A KPI monitoring pipeline for average call duration was set up to automate this procedure for future incidents.

\section{Lessons Learnt}
\label{s:lessons}
Through designing and working with \libname we have learned several lessons that are shared in this section.  These include lessons about tackling the problem and using the tool.

\subsection{Not Enough Data}
Since KPI metrics are critical to understanding business health, they are often monitored outside of an anomaly detection workflow.  There is a tendency to want to explain all movements in the metrics. Investigating all changes outside of an automated workflow quickly becomes a drain on resources. As a result, two tenets that our data science team maintain are to be clear about defining what constitutes a regression in a metric and determining the volume of data needed for investigations.
 We work collaboratively with engineering teams to determine what threshold for metrics regressions are important.  This is based on the expected variation of the metric, along with the business impact.  Based on the size of the regression, the volume of calls needed to detect those changes is determined.  This means that if a metric regresses on a slice of data with a volume of calls below the threshold we say that there is not enough data to follow up.  
 When \libname is run with insufficient volumes of data, the statistics test in the initial metric comparison determines the metric to be within an expected range and the feature ranking component of the algorithm returns an empty list.  

The tool and the deployment workflow are not guaranteed to catch all regressions in the service. Recently we saw an incident of this form: there was a faulty device in a specific set of carriers for our PSTN service that was in a small volume of data in the call corridor. The movement in the KPI metric for this was within the range of expected behavior and was not detected by our time series anomaly detector.  \libname also did not provide insight on this data for the same reason of not having a large enough call volume.  In practice, it is important for service owners to have multiple tools to monitor for regressions and have manual overrides based on strong domain knowledge.

\subsection{Feature Selection}
For our scenarios, we collect thousands of points of raw and processed telemetry for each call.  An initial impulse was to include everything into the hypothesis features.  This was shown to have several problems: 
\begin{itemize}
    \item Some features are derived from the metric resulting in variable leakage \cite{guo2010identifying}. These features do not provide any actionable information and must be removed.
    \item Features that won't make sense to explain the metric.   For example, telemetry collected post-call setup should not be used to describe an increase in call setup failures. 
    \item A large number of correlated features leads to results that are difficult to interpret. A small amount of investment to prune out strongly correlated feature sets, while keeping a representative feature within each set served us well in the long run. This helped engineers cut through the noise and consume the results easily.
\end{itemize}
We have found that having a collaborative process of feature selection with the engineering teams using \libname has resulted in the best performance of the algorithm.  From the data science side, we used historical data to filter the potential list of hypothesis features down to a manageable number.  The engineering teams further curated this list with respect to their domain knowledge. Based on our experience with \libname we find having around 200 hypothesis columns provides a good balance between giving actionable insights while still being accurate.

\subsection{Multivariate Feature Ranking}
When creating the tool we held a philosophy of keeping the results as simple as possible so that engineers without a data science background could quickly process the results.  This led to a univariate feature ranker which has shown success in providing a direction for a follow-up investigation.   

There have been past investigations when surfacing a multivariate path that explained the metric increase would have enabled a quicker investigation.  In the case study for average call duration detailed in Section \ref{s:deply_nonbinary}, having surfaced the primary tenants impacted in conjunction with the area codes would have shortened the investigation time. For this particular incident, we approximate the time savings would have been less than 1 hour.  Given the volume of investigations and the clarity this additional information can provide, it is an area worth investing in.  

Developing a multivariate feature ranker can be done by creating new features.  To create a bi-variate feature ranking a new feature is constructed based on considering two features simultaneously.  This increases the number of features quadratically. We found that blindly considering all combinations does more harm than good as it overwhelms engineers with more information than they can process. Striking a good balance between considering multiple features and containing the noise level requires an efficient pruning algorithm that minimizes the number of combinations considered. 

\section{Conclusions and Future Work}
\label{s:conc}
This paper introduces \libname, a library that tackles the challenge of diagnosing metric alerts generated by services running at scale. We built and deployed \libname for filtering out anomalies due to population biases and providing engineers with
actionable root-cause information for genuine ones. \libname is running as part of a live system deployed to help with the diagnosis of services within Microsoft's Skype and Teams engineering services. Based on our experience, we found anomaly detectors fired as much as $90\%$ of the time due to changes in population demographics. This simple result of piping an anomaly detector with a bias checker results in a $95\%$ savings of the engineering team's time allocated towards investigations. To the best of our knowledge, this is the first work that proposes checking for alerts against changes in population. This phased approach of checking for changes in user population followed by a ranking of service features serves two benefits. First, it provides product owners with key insights about demographics changes of their application, and second, it identifies opportunities for service owners to improve their engineering system. We make \libname available to practitioners by making it available for download and present multiple case studies to help other practitioners learn from our experience. 
In the next phase of our work, we plan to focus on expanding support for continuous metrics, perform feature ranking using multi-variate features,
and introduce feature clustering to tackle the problem of multicollinearity in feature ranking.

In conclusion, we emphasize that \libname is a novel approach designed for engineering teams to manage the deluge of metric alerts, enabling them to spend less time in diagnosing metric regressions, and more time on building exciting features.

\bibliographystyle{ACM-Reference-Format}
\bibliography{mct}
\end{document}